# The Social Value of Dark Energy

Avner Offer (Corresponding author)

All Souls College, University of Oxford

Avner.offer@all-souls.ox.ac.uk

ORCID 0000-0003-4762-0277

Ofer Lahav

University College London

o.lahav@ucl.ac.uk

ORCID 0000-0002-1134-9035

**Conflict of interest**

Avner Offer declares that he has no conflict of interest. Ofer Lahav declares that he has no conflict of interest.

**Acknowledgments:**

OL thanks All Souls College, Oxford for a Visiting Fellowship in 2021, where this collaboration began. Urvi Khaitan has contributed capable research. Grateful thanks to Jeremy Butterfield, Lucy Calder, Tom Diehl, Brenna Flaugher, Massimo Florio, Josh Frieman, Rich Kron and Alexander MacDonald for their helpful comments.

(16 May 2023)

# The Social Value of Dark Energy

**Abstract.** Astrophysics is a social enterprise exemplified here by the Dark Energy Survey (DES) which completed its fieldwork in 2019 after 16 years of preparation and observation, while data analysis continues. Society funds astrophysics on a grand scale. For human capital and for governance the discipline draws on a self-governing 'republic of science', while the funds were provided by philanthropists in the past, and by governments today. The benefits accrue initially to scientists themselves, in the form of a rewarding vocation. For the social benefit it is tempting to apply formal cost benefit analysis, but that approach ignores the option value of science and imposes questionable assumptions from welfare economics. Astrophysics generates some useful spinoffs, offers attractive careers, appeals to the popular imagination, speaks to metaphysical cravings and constitutes a good in itself. The rise of AI also suggests a role in exploring future habitats for intelligence and cognition.
Keywords: astrophysics; dark energy; cost-benefit analysis; social value

**Introduction**

Astrophysics is an exact science and also a social enterprise. As a discipline it cannot be understood with the same accuracy as its findings. In astrophysics observations are precise and align tightly with theory. Observer and observed are separated. Data are well-behaved and replicate consistently. Humanity and its planetary environment are different: unlike the heavenly bodies, earthly ones often have a wilful intentionality. Observed and observer overlap, outcomes cannot be foreseen, findings are difficult to replicate, multiple perspectives exist. Earth science, biology, meteorology, social science, and the humanities all have temporal discontinuities and explanatory shifts over time.

  Astrophysics creates value for society which is what allows it to exist. These values are diffuse and not easy to add up. We approach them by means of a single middle-sized astrophysical project, the Dark Energy Survey (DES). Based on our own insider knowledge of the project, the US funding agencies and universities combined with international agencies, institutes and universities have contributed a total of about $100 million. Researchers are still being paid to process and analyse the data, directly by their home institutions, or via grants awarded to individual investigators. .

DES promised an incremental contribution incremental to cosmological understanding. Such projects provide large rewards to those who work in them, material and cognitive, emotional and social. The public agencies that underwrote the cost considered it value for money. Why are those agencies there, and what do they do? Such funding can be controversial, and is challenged from both the right and the left. Cost-benefit analysis (CBA) is a formal procedure of social valuation. It fails to capture the benefits that motivate funding agencies, and misses the element of speculative promise in scientific enquiry. Several other sources of value are estimated here, some of them crisply and others more tentatively. Are the social benefits of astrophysics an intrinsic good in itself, or is their value practical? The activity has some valuable spinoffs, but our initial inclination was to regard the quest as an end in itself. Somewhat to our own surprise, the conclusion points towards a role for astrophysics in the unfolding of post-human intelligence, which raises questions about human existence itself.

Human activity is approached in terms of cause and effect, while the universe is often seen as having no purpose or reason: 'The more the universe seems comprehensible, the more it also seems pointless' (Weinberg 1978: 149). But humanity and the heavens are not separate: they are linked by the act of observation. There is a view (the 'anthropic principle') that for the universe to be observed, some cosmological preconditions are needed to make humans possible, but most of its proponents regard these conditions as arising by chance (Barrow and Tipler 1986; Bostrom 2002).

In General Relativity (Albert Einstein's cosmological model) space-time is configured by the mass of heavenly bodies. But visible matter falls short of explaining the universe's observed parameters, and especially its accelerating expansion. To fill this gap, two entities have been postulated, namely 'dark matter' and 'dark energy', dark because they cannot be observed directly and are poorly understood. Dark matter brings the observed cosmic structure into agreement with theory, and dark energy reconciles theory with accelerated cosmic expansion. The acceleration is observed 'as if' driven by 'energy', the nature of which is unknown (it might be due to quantum fluctuations in the vacuum). Several collaborative projects have set out to obtain measurable parameters for these entities. One of these, the Dark Energy Survey (DES), is used here to understand why such projects are undertaken, and what is required for them to take place (an overview in Lahav et al. 2021).

**The Dark Energy Survey**

Surveying in astronomy is divided into spectroscopic and imaging. Spectroscopic surveys produce accurate redshifts (approximate distances) to galaxies, but are demanding in terms of observation time. Multi-band photometric surveys can measure far more galaxies, but the deduced 'photometric redshifts' are less accurate and precise than spectroscopic ones. Space-based missions produce better images than telescopes on the ground, as the observations are done above the Earth's atmosphere.

Back in 2006, the U.S. Dark Energy Task Force (DETF) report classified Dark Energy surveys into stages: Stage II projects were on-going at that time; Stage III were near-future, intermediate-scale projects; and Stage IV were larger-scale projects in the longer-term future. Dark Energy has a high scientific priority: about a dozen projects are in progress in several countries, among them DESI, Rubin-LSST, Euclid, and Roman-WFIRST. Our focus is the Dark Energy Survey (henceforth DES), a stage III project (Dark Energy Survey Collaboration 2016: Lahav et al. 2021). Its observations are already complete (as of 2019), results have been published in over 400 articles, and the final analysis is still underway.

We have an insider's view, as one of the authors (OL) has been part of the project since its early days in 2004, in particular as co-chair of its Science Committee until 2016, and later as chair of its Advisory Board. DES is an imaging survey of one eighth of the sky, utilising a dedicated 570 mega-pixel camera (called DECam) on the 4m Blanco telescope in Chile. Redshifts (approximated distances) of the galaxies were obtained from a sample of thousands of Supernovae, and from multi-band photometry, which made it possible to produce a three-dimensional map of 300 million galaxies.

Einstein (1917) hypothesized a 'Cosmological Constant' (labeled by the Greek letter $\Lambda$ (lambda)) that corresponds to a model universe with density that is constant in time and in space. One of Einstein's motivations was to have a 'static' (neither expanding nor contracting) universe. When Hubble's observations showed that the universe was actually expanding, Einstein allegedly referred to $\Lambda$ as the 'biggest blunder of his life'. But $\Lambda$ can also take other values, that give e.g. an expanding and even an accelerating universe, as current observations suggest. Furthermore, the concept of $\Lambda$ can also be generalized to 'Dark Energy'. The main goal of DES (and other similar surveys) is to determine the constant Dark Energy 'equation of state' w (the ratio of pressure to density) and other key cosmological parameters to high precision (at the level of a few percent). The value of w that allows density to be constant at any cosmic time is exactly w=-1 (this implies a somewhat non-intuitive concept of negative pressure!). The concept of Dark Energy however allows for other possible values for w, including a w that varies with time. The big question is if the Dark Energy equation of state w=-1 exactly, or is something different.

DES measures in seven different ways, having developed some new ones since it was started: weak gravitational lensing, galaxy clustering, Baryonic Acoustic Oscillations, cluster abundance, standard sirens, Supernovea type Ia and strongly-lensed transients. So far, based on analysis of about half the data, the headline result from DES galaxy clustering and weak gravitational lensing, combined with the Planck Cosmic Microwave Background experiment and other probes, is that w= -1.03 +- 0.03 (DES collaboration 2021), consistent with a Cosmological Constant, w=-1, as explained above. In other words, the combination of three sets of elaborate empirical surveys using different methods have converged to produce a result consistent with the unique cosmic parameter of w=-1. This value has been postulated theoretically and is a very small target to hit empirically.

This is not yet the final word.  It is difficult to know  when to stop, and further massive experiments continue (Lahav and Silk 2021).

The survey had its first light in September 2012 and started observations in August 2013. Observations were completed in 2019 over 758 nights spread over six years. DES  imaged a depth of approximately 24th magnitude. This is about 15.8 million times fainter than the dimmest star that can be seen with the naked eye. On a typical clear observing night, the camera took about 200 images;  each DECam digital image is a gigabyte in size, so DES typically collected about 200 gigabytes of  processed data per night and 5-10 times more of raw data,  which is 'big data' for an astronomy experiment. The DES data are publicly available and can be downloaded from the internet. DES scientists also work closely with data sets and scientists from other surveys and observatories.

**Sufficient and necessary conditions**

The Dark Energy Survey has spent about $100 million in direct costs alone. By the standards of the social sciences (though not of astronomy) this is a very large sum. Raising it may be regarded as a necessary condition for the project to exist. Other conditions are a pre-existing research infrastructure and an accommodating social environment. These prerequisites have been centuries in the making.

The universe emerged almost 14 billion years ago. Astrophysics in its modern form goes back about five centuries. The period between the $16^{th}$ and the early $20^{th}$ century has been called 'the long space age' (Macdonald 2017), the era of princes and patrons. Astronomical theorising and observation were first endowed by dynastic elites in Europe, later joined by wealthy individuals in North America. In this respect astronomy was somewhat like opera: prestigious, expensive, and self-indulgent (McAndrew 2006). During these centuries the main drive came from enterprising investigators supported by wealthy patrons. Sometimes the value promised was practical, affecting for example the problem of ocean navigation in the $17^{th}$ and $18^{th}$ centuries, while at other times the inducements verged on deception (Macdonald 2017).

In early modern Europe, from the 16th century to the 18th, an international scholarly network existed in the form of a putative 'republic of science', with local nodes in learned societies, royal academies and universities (Polanyi 1962; David 2014). The sociologist Robert K. Merton famously described its emerging norms with the acronym CUDOS, standing for 'communism' (in the sense of common ownership, of findings open to all), universalism (disregard of social background and religious orientation), disinterestedness (i.e. not standing to profit from outcomes), and organized scepticism (Merton 1942: 118-126). Science was 'open' in the sense that findings were freely shared and published for all to read. Scholars were evaluated by peers, not according to social status or wealth but on what they added to knowledge. Investigation went wherever it led and scholars could speak openly subject to critical scrutiny only by their peers. For the chemist and philosopher Michael Polanyi, who originated the term 'republic of science', the crucial attribute was self-government (hence a 'republic'), with authority arising from the informed consensus of qualified experts. Authority was brought to bear by means of admission, appointment, promotion, publication, citation, and informal consultation and mentorship (Polanyi 1962).

These implicit norms were explicitly stated in response to Nazi and Soviet attempts to align scientific research with ideological imperatives, and the persecution by these regimes of non-conforming scientists. Merton and Polanyi both published their paradigms of open science in the early 1940s and both of them were of Central or Eastern European Jewish lineage. At the time, with such social origins and the requisite talent, science presented an attractive career opportunity. 26 percent of Nobel prize winners in physics between 1901 and 2000 were of Jewish heritage (Shalev 2002: 89-90, tables 21A-B). The concept of a Cosmological Constant itself (and its later extension to dark energy) originates in that nexus: Albert Einstein was of German Jewish background, and after rising to the pinnacle of German science, he had to escape overseas when the Nazis came to power.

In early modern Europe both patrons and scientists engaged in reputational competition, albeit a different one for each group. Within a politically fragmented Europe elites still had a good deal in common: Christianity, universities, a Latin lingua franca, dissemination by print and postal services. Kings, princes, nobles and prelates competed for reputation and prestige. Scientists also competed, for renown among their peers.

Elite competition and scientific investigation were co-dependent. In science, reputation and rewards are achieved by priority of publication. Scientific findings make a good competitive signal which is difficult to make and difficult to fake. Scientific achievement embellished reputations for elite patrons, but only scientists could validate it. Validation was obtained through the CUDOS procedures of the republic of science, which later discharged the same function for the nation-states and great power empires of the 19$^{th}$ and 20$^{th}$ centuries, and continue to do so today (David 2014).

During the nineteenth century German universities became scientific powerhouses and provided models for institutions elsewhere, especially the North American research university. Despite being embedded in illiberal regimes and the civil service status of their professors, German universities committed themselves to the growth of knowledge: to the principles of academic freedom, to integrating teaching and research, and to open-ended inquiry (Menand et al. 2017: Introduction). That they operated in an oppressive political environment may account for the insistence that knowledge should be apolitical and value-free (Weber 2015). Indeed, this may have already been necessary for the emergence of impartiality in the previous era of noble and philanthropic patronage.

During the interwar years the market dominance of several large corporations (e.g. Du Pont and AT&T in the United States, Phillips in the Netherlands) gave their research labs the leeway to pursue basic science. Large foundations of corporate origin (Carnegie, Ford, Rockefeller) endowed University research and sometimes set up their own academic institutions (e.g. the Russel Sage Foundation, Rockefeller University).

A new phase of secretive 'Big Science' began with the Manhattan Project in the Second World War. Academic scientists in industrial-scale government research laboratories created nuclear weapons and led the way into the post-war model of science. Cold War competition advanced the frontiers of research. Military funding flowed into the universities. Government expanded its own research laboratories, especially in high-energy and particle physics. NASA managed the moon shots of the 1970s. The National Institutes of Health drove biological and medical research, while the National Science Foundation funded basic research in both natural and social sciences (Westwick 2003; Mirowski 2011: ch. 3; US National Science Board 2020).

Government support for research and development in the United States peaked in the 1960s at about two thirds of total expenditure. From the mid-1960s business-funded research began to rise and the two curves intersected in 1980. By 2017 the proportions had been more than reversed, with business at 70 percent, and government at 22 (US National Science Board 2020: fig. 4-4, 18). In the new regime open science was replaced by intellectual property, only to be disclosed after patents were secured. In 1980 the Bayh-Dole Act allowed universities to appropriate, patent and profit from government-funded research. This reduced the incentive to publish. Universities allowed individual investigators to file patents and exploit them commercially (Sampat 2006). A few top institutions harvested large windfalls. For most research universities however commercialisation did not pay off but it stifled the sociable conventions of open science, even within single academic departments (David 2007: 269; Sampat 2006). Two Federal Technology Transfer Acts (in 1980 and 1986) encouraged commercial exploitation for national lab research. In the 1980s the US government began to fund leading-edge commercial research directly, allowing the companies involved to keep the intellectual property created (Cohen and Noll 1994). Business needed to get to market first so research and development had be confidential (e.g Kidder 1981). But long-term risky open-ended research is difficult to undertake in competitive markets because returns are uncertain (Offer 2022). In high-energy physics and astronomy there is little prospect of profit for business apart from occasional spinoffs. This is one reason why United States and other nations continue to carry out so much research in government institutions.

The shift from public to private was largely driven by internal developments within science itself. Pharmaceuticals and healthcare found ready markets. Falling prices drove rapid diffusion in information technology. While governments paid the development cost of weapons, the profits were made in producing them (Cohen and Noll 1994; US National Science Board 2020). In the Cold War years the republic of science was a continent of open research. By the early 21st-century it had broken up to become an island archipelago in a proprietorial sea (Mirowski 2011: pt. II).

**Towards the Dark Energy Survey**

The Dark Energy Survey originated in discussions between scientists at Fermilab (one of the National Laboratories), the University of Chicago, and the University of Illinois at Urbana-Champaign. Following the Supernova results of 1998-9, it began to take form in the early 2000s in academic conferences and informal consultations. Scientific challenges were identified, their scale estimated, logistics worked out (Lahav et al. 2021). The project was scientifically scoped and then costed, which made it possible to identify funding options. Fermilab, located west of Chicago, is a freestanding campus engaging mostly in particle physics experiments, and employing 1,750 people today. For scientists working in such a setting a research grant in the tens of millions was not difficult to envisage.

The patron-client relation in science gives rise to a principal-agent problem (the term 'client' is used here in the sense of a dependant subordinate, not that of a customer). The principal pays for the project, but as a patron they are poorly equipped to decide whether projects are good value, all the more so in a quest for the unknown (Guston 2000). In comparison with business and universities, government agencies commission a good deal of scientific and technical advice to validate their funding decisions. In consequence government funding routinely imposes onerous cost accounting, 'far more elaborate, costly and inflexible than the monitoring systems that private organizations employ for their own research' (Cohen and Noll 1994: 74). These routines also impose an orderly progression on funding applications. The American initiators of the DES mapped out a likely sequence. Experts were identified and invited. An international network of specialists drafted the detailed submissions required, funded partly by the participants' own salaries and research budgets. Working groups hammered out the detail in a partnership between the republic of science and its government patrons.

The main cost elements were observational hardware, telescope time, and computer facilities, as well as the requisite technical staff. Only deep pockets could undertake such large up-front expenditures. On the hardware side a new digital camera had to be designed and built for the telescope in Chile. A grant of £1.7 million was obtained by a British group, which provided credibility for the main funding application of $35.1 million to the United States Department of Energy. National Science Foundation participation was critical to the project and its approval. NSF paid for the Telescope upgrades and the data management system. The application went through an exhaustive sequence of evaluation until the grant was approved, incorporated into the United States budget, and signed into law by President G.W. Bush in 2008 (Lahav et al. 2021: 49-51, 397-399)

  A formal management structure for the project was then set up and the project unfolded in sequence through the stages envisaged: recruitment of staff and allocation of tasks, followed by commissioning, construction, installation, and testing. An observational effort on such a scale generates colossal troves of data which require a computing infrastructure (software and hardware) to acquire, analyse, store, and disseminate. From then on everything went approximately according to plan: observation, data collection, analysis, dissemination, and publication.

  The remit of the Department of Energy was mostly nuclear: In addition to nuclear weapons, it funds costly high-energy research in particle physics (and also promotes applied energy innovation, both public and private). This cabinet-level Department is also a scientific and technological component of the American international order. As in NATO, the United States is the leader and driver. An 'economic theory of alliances' postulates that junior partners get away with less than their proportional contribution, confident that the senior ally will provide the funds regardless (Olson and Zeckhauser 1966; Sandler and Hartley 2001). Open science is a public good whose findings cannot be withheld. Hence junior partners can participate at low cost to themselves. Scientists took advantage of opportunities created by international competition and national defence, while the funders hoped for scientific and technological windfalls which might serve their own purposes (MacDonald 2017: ch. 3; Hughes 2002: 159). Like astrophysics, national defence is also a public good exempted from the calculus of commercial profit and with only loose constraints on expenditure, which makes it a good partner for expensive science.

The European Union also undertakes big science. In the CERN Large Hadron Collider particle accelerator it took on a challenge abandoned earlier by the United States for reasons of cost (Riordan et al. 2015). It is currently constructing an 'Extremely Large Telescope' in Chile which is claimed to be more powerful and more advanced than its two American counterparts (Witze 2021). The European Union is not a military power but it is a distinctive political cluster which (to be consistent with our line of argument) points towards international prestige competition as a driver of large-scale funding of basic science, in line with the patron model of early modern science.

Unlike the soft human sciences, the Dark Energy Survey requires an array of hard failsafe skills (Pye 1964). The project falls within the bounds of established theory, and is designed to corroborate or falsify it. It is technically challenging but deliberately confined within the scope of existing technology and skills. DES was able to deliver on everything it promised, construction and observation were completed on time and on budget, while scientific analyses are still going on, probably for another couple of years. In that respect it is more like research and development than open-ended science. Hence its highly structured allocation of resources, rather like the design of a new car or aircraft, is intended to resolve a pre-defined development problem with a high expectation of success. There is still room for unexpected findings: In the case of DES, a falsification would open up new scientific horizons.

Other imperial ventures of the same years, e.g. the American wars in Iraq and Afghanistan, may also be regarded as not-for-profit long-term open-ended 'public good' projects. These speculative great power enterprises have gone badly wrong. In contrast, the Dark Energy Survey objective is firmly under control. Hence the process of approval bears the hallmark of top-down corporate resource allocation rather than democratic 'republic of science' procedures. The iron law of major projects is 'over budget, over time, under benefits, over and over again' (Flyvbjerg 2017: 12). The reasons are incomplete knowledge and strategic behaviour (Offer 2022: 18-21). The James

Webb Space Telescope (JWST), a US-led venture, was launched in 2021 at a total cost of $10bn, almost seven times the initial estimate of twenty years before (Billings 2021). In contrast, the DES has delivered almost on time and on budget in a tribute to its spirit of prudence and willing cooperation.

In astrophysical terms the DES belongs to a class of medium scale projects, costing in the tens of millions of dollars. The United States commits periodically to astronomical projects which are larger by orders of magnitude, like the Hubble Space Telescope and its successor the JWST. These commitments are updated every 10 years or so and can take decades to complete (Billings 2021). Several prospective dark energy studies are going to be an order of magnitude larger than DES. For all its remarkable ambition, complexity, and achievement, DES is something of a technosphere commonplace, about the selling price of a low-end airliner, and taking about the same time to develop, at a small fraction of the cost.

**University home bases**

DES is a meritocracy. Some four hundred scientists working in more than 15 countries collaborate to carry out a highly complex sequential process. Their intrinsic abilities and acquired skills are scarce among the general population but are abundantly available for projects such as DES. Astrophysics is a mature science with a solid presence in research universities. The personal entry hurdle is demanding and qualified staff are highly capable. Academic staff are a free resource for projects such as DES, since they continue to be paid by their home institutions. Here we have chosen, as a rough measure of human capital available to DES, to take a sample of the 170 authors of a recent research paper (Dark Energy Survey collaboration 2021). For the 153 (90%) for which this information was available, on average each scientist had a stock of 19.3 years of experience between first degree graduation and the submission of the paper in May 2021. Multiplied by the 400 scientists engaged on the project (and assuming, not securely, that non-authors were not systematically different), that makes up some 7720 years of scientific experience (a.k.a. human capital) brought to bear on dark energy in this single project by 2021. Precision is not significant here. The number merely indicates that DES has the benefit of thousands of unpaid-for years of experience. For access to this human capital DES had only to cover, and that only partially, subsistence, travel, and research expenses. The scale of astrophysical research exceeds the capacity of any single university. These institutions delegate their staff, and in return they get to harvest some of the credit.

The currency of the republic of science is reputation. Academic scientists get ahead by building up credit for the co-authorship of academic papers which make an impact (Calder and Holbraad 2021). Some work in astrophysics requires participation in large research teams. This gives rise to a benign trade-off: scientists are keen to be chosen and project leaders have talent to choose from. In large research projects postdoctoral researchers can earn the credits they require to move into permanent positions. Much of the work in DES, however, does not consist of publishable science, but of skilled and even creative support work, for example in data acquisition, analysis and dissemination. Such work is less than fully acknowledged by publication. In astronomy postgraduates greatly outnumber the permanent posts available, which may represent loss or even tragedy for those competing, but helps to ensure a high entry standard. Doctoral and subsequent training in astrophysics takes about a decade, which is lost to aspirants if a post is not secured. If postdoctoral scientists have no insight into their own personal competitive chances, then part of their incentive (over and above their low pay) is the lottery value of a permanent academic post, an incentive mechanism already described by Adam Smith for professions like the military and the law where many are called and few are chosen (Smith 1776/1976: Bk I, ch. X). Their personal loss is not however one for society. Scientists qualified up to doctoral levels find ready employment in many activities, and it appears that most do not pursue scientific careers (Florio 2019: ch. 4). There are therefore programmes which prepare students for such dual careers, e.g. the Centre for Doctoral Training in Data Intensive Science (CDT-DIS) programme at University College London (https://www.hep.ucl.ac.uk/cdt-dis/). In the DES publication early career researchers (up to 10 years of experience) constituted by far the largest group, followed by those with 10 to 15 years of experience, some of whom will have entered the project as doctoral students or post-docs. This is consistent with the impression from Calder and Holbraad (2021) that younger researchers are highly represented in the project (figure 1).

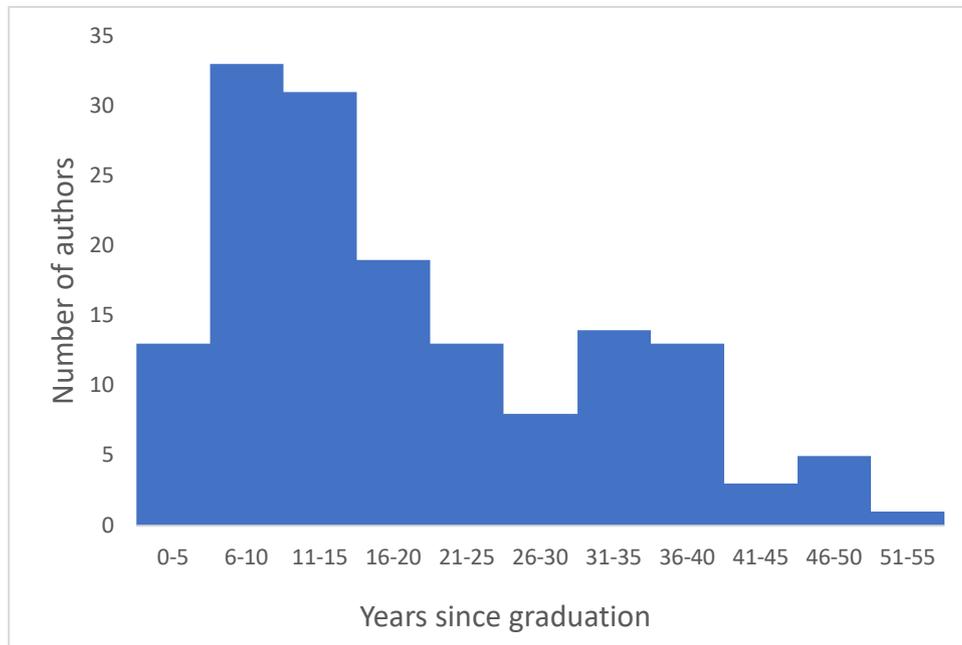

Figure 1. Years elapsed since Graduation (BSc) of authors and publication of Dark Energy Survey collaboration (2021). Data on 153 out of 170 authors. The mean is 19.3 years.

Universities form the permanent infrastructure of science. They maintain and renew the stock of formal and tacit knowledge, they identify talent and train it. They are also the nodes of international research networks, made up of permanent and ad hoc collaborations, and provide the channels through which scientific conversation diffuses. In the DES research article mentioned above, 97 of the 170 authors are from the United States, 24 from the UK, 19 from Spain, 15 from Brazil, 6 from Germany, four from Australia, two each from Chile and Switzerland, and one each from Canada, France, India, Italy, Japan, Norway, and Portugal (multi-national affiliations are credited to all the countries involved). University collaborations also underpin the specialist operation of free-standing scientific installations, the fixed capital, the optical, radio, and satellite telescopes constructed in remote locations, and the computing capacities that support them (Florio 2019: Introduction).

**Science under fire**

Science has come under attack from two forces of modernity, corporate business and populist democracy. Loathing of science has increased under the market liberalism which came into power in the West around 1980 (Mirowski 2020). It embodies two different attitudes: the profit-seeking thrust of business and finance, and the anti-elitism of the common man and woman.

One purpose of market liberalism was to transform the public sector into a source of profits, rents, and interest payments (Offer 2022: chs. 2-3). In the United States the Reagan regime soon allowed universities to profit from work undertaken with government funds. The payoffs were spectacular in IT and bio-medicine, mostly enriching the scientists themselves rather their institutions, thus creating incentives to privatise the findings of open science. In consequence published research in the United States began to decline while a development work previously done in universities was increasingly outsourced to cheaper locations overseas, like manufacturing in the same period (Mirowski 2011: ch. 6). China has already overtaken USA scientific research in both quantity and in quality (Wagner et al. 2022; Lu 2022).

From the other side democratic critics of science began to demand that it should be more responsive to popular needs. Philosophy of science used to be concerned with the validity of scientific knowledge. Starting around 1980 a sceptical 'progressive' strand has worked to show that a good deal of scientific discourse is constructed to underpin privilege, not least the social privilege of scientists themselves. In the so-called 'science wars' within a new discipline of Science and Technology Studies (STS), radical social constructivists dismiss the validity of science altogether. More moderate critics acknowledge the argument but continue to uphold scientific norms as embodying a valid if not entirely disinterested pursuit of truth (Mirowski 2020).

A pseudo-democratic attack comes from the right. The market liberal revolution advocates competitive markets and top-down market power at the same time, although these two objectives are not consistent with each other. It has little patience with social planning, the policy equivalent of scientific method. Let markets rip and business grab, private gain before the common good, since market competition maximises the total well-being. 'We've had enough of experts', declared the Conservative politician Michael Gove during the Brexit campaign of 2016 in Britain (Mance 2016).

This disdain for science has brought together the anti-Enlightenment tradition of conservatism and the anti-intellectual bias of much of the population into a political alliance in many countries. One of its attributes is resistance to scientific evidence and the policies on which it is based, most notably in climate change denial and vaccine resistance. One might surmise uncharitably that both the social constructionist view of science, and the disdain for scientific rationality, may reflect a resentment of the social standing conferred on those who master science.

Astrophysics is largely immune to these controversies. It is elitist without and meritocratic within. It retains some Cold War attributes of government funded 'Big Science'. Like the military projects from which it remotely originates and which it partly mimics, astrophysics benefits from widespread social support perhaps out of a vague notion that it is protective. There is certain amount of self-indulgent rocket and satellite development among the very wealthy and some commercially viable low orbit satellite enterprise but astrophysics as science more generally is difficult to commoditise: It has no consumer product, no corporate incentives, no start-up opportunities (some technological by-products are treated separately below). The knowledge required is beyond the reach of Everyman. As a discipline it is held in reverence and awe.

**The limits of privatisation**

Markets are the norm in economics and in much of political and media discourse. From that point of view astrophysical knowledge is an anomaly, an intangible public good that does not arise out of commercial exchange. In truth however the astrophysical enterprise is closer to the real existing norms of the economy than competitive (so-called 'free') markets. Competitive markets are limited in scope by the requirement for business to break even within a circumscribed period of time, the outer bound of which is defined by the interest rate. This is the 'payback period': the higher the rate, the shorter the wait (Offer 2022: 11-18). Business typically has to make a profit in less time than that. Any enterprise which cannot break even on its own within the payback period requires a different approach: public enterprise, not-for-profit, or philanthropy. The payback time boundary can be overridden by government assigning a 'franchise', defined here as a private revenue flow with some protection from competition by means of pricing power, long duration and low variance. Such revenue flows are available to 'natural monopoly' network utilities, electricity, gas, water and landline telephones, or strong commercial brands supported by advertising or intellectual property rights. Governments support franchise operations in a variety of ways. Hence a great deal of business enterprise has little to do with competitive markets and is undertaken at the pleasure of the state. Commercial banking, whose lending rate defines the payback boundary, is itself a franchise, underpinned by central banking (Offer 2022: 21-22).

Many valuable activities cannot pay off within the payback boundary, if they ever pay off financially at all: health, education, science, art, universities; nuclear power; defence and war; railway projects like the UK Crossrail and HS2; motorways (freeways, highways); urban and national parks; museums, libraries, symphony orchestras, and opera; space exploration; mitigating climate change. Such activities, with long durations and fixed capital locked in, cannot be undertaken for profit by business alone. To carry them out, the temporal constraints of commercial credit need to be overcome.

These activities all take time to deliver. Nothing is longer-lived than the universe, and its study falls in this respect into the same category as other long-lasting goods, such as landscape, religion, the arts, major infrastructure, life-cycle social services, environmental protection and arresting climate change, none of which can turn a profit in competitive markets (Offer 2022). National defence is also good of this kind, and the skills and methods of astrophysics benefit from an association with those of the military. These long-term attributes place astrophysics within the public sector.

**Social and private payoffs: Cost benefit analysis**

Cost-benefit analysis is an attempt to apply the precision of science to the decision of whether to undertake it. A business project is worth pursuing if it can make a profit. For the public sector, where there is no profit, cost-benefit analysis (CBA) provides an alternative hurdle, albeit one that is not easy to construct. In CBA a public sector project is socially worthwhile if the sum of its benefits adds up to more than its total costs, measured in a uniform unit of account (e.g. dollars) and discounted over time. Benefits and costs which have no market value (or whose social value differs from their market value) are imputed a 'shadow' price. This is laborious and imprecise but has been applied to thousands of infrastructure projects and helps to prioritise public spending. In science, cumulative outlays are not too difficult to estimate. The focus is therefore on estimating benefits. We follow Florio (2019) who has applied the method to Scientific Research Infrastructures with special reference to the Large Hadron Collider at CERN. This is a pioneering and accomplished study by an experienced investigator.

In this approach the prime beneficiaries of the DES are the scientists themselves. In cost-benefit accounting the social value of public expenditure (as in national accounting) is taken to be the same as its cost, so the value of scientific work is captured by staff and equipment costs alone, with no presumption of social value-added. The future value of scientific discovery could turn out to be both positive and negative, and is therefore conservatively set to zero and disregarded (Ibid. 40, 42). There is however a small net benefit which arises from the downstream value of research as measured by citations. This value is modest since the benefit is estimated as the time cost of reading scientific publications.

Research Infrastructures and the projects they host employ many doctoral and postdoctoral early career researchers (ECRs). For example, in the DES author sample used above, some 30 percent are ten years or less since first degree graduation and others who were already a few years over this limit (a large category, see figure 1) had spent time in the project before they crossed it. Over and above the value of their contribution (measured by its cost) their work also constitutes training which may be regarded as investment in the formation of human capital. Its net social value can be estimated in a CBA by comparing the positive surplus of lifetime earnings of ECRs over those of similar qualifications who did not participate in such large projects. This premium can be substantial and is calculated to be the largest single CBA net benefit of the CERN Large Hadron Collider, though its actual magnitude has to be speculative since much of it falls in the future. It is possible, however, that the two groups are not comparable, and that CERN participants are selected for their superior endowments, in which case this benefit may be spurious in whole or in part. Many of these researchers leave science subsequently but their human capital appears to attract an income premium in other occupations. This is a private benefit. Value-added for society however is only certain under neoclassical labour market assumptions in which wages are assumed to be equal to their marginal (social) value. Much demanding work in finance or defence is highly paid but of questionable social value, though that is not a judgment that a CBA will typically make.

Knowledge spillovers from large projects add value for direct suppliers and other firms and organisations. In the DES the new hardware, the digital camera, was fabricated in a collaboration between DES scientists and private firms. An instance of innovation value-added is the European Union's Copernicus Sentinel satellites which provide a global, continuous, autonomous, high quality, wide range Earth observation capacity, and make their data available free of charge (Florio 2019: 209-211). The global positioning services (GPS) and the invention of WiFi by radio astronomers may also be considered as offshoots of astrophysical research. But like the future value of scientific discovery, it is not reckoned to be an expected benefit of CBA, though it is reasonable to assume that some such benefit will ensue. As in the case of knowledge value-added, not all such technologies are necessarily beneficial.

Research infrastructure projects create large flows of data. As mentioned above DES has generated several Gigabytes of raw data per night, and the next generation of surveys will generate at least an order of magnitude more. Work on data analysis began almost ten years before 'first light' (Lahav et al. 2021: chs. 6-7). Ultimately it drew on the resources of the US National Centre for Supercomputing Applications. Managing, analysing, storing and making these data flows available required innovative coding. In common with most other major RI projects of the last two decades the DES complies with 'republic of science' norms and makes its data available online free of charge, although it is not clear how this should be valued in CBA accounting conventions. A large number of articles using DES data have originated outside the project, and published by researchers with no formal connections to it. This appears to be a measurable benefit over and above cost, though one that is difficult to quantify except in publication numbers and their measured impact.

There is a popular demand for astrophysical psychic stimulation. Astrophysical findings make media headlines. Large numbers of people are reached through science fiction by means of film, television, and print. Popular science is available in articles and books. Sensory uplift is generated in video documentaries, public lectures, museum displays and computer games. The rudiments of astrophysics are taught in school. The time and money spent can be estimated in principle to measure the net value added.

**Cost-benefit queried**

The question is not whether scientific research gives rise to economic benefits, but whether economic benefits capture all the welfare that science produces. 'CBA will never be the core argument to justify the investment of taxpayer money in science' (Florio 2019: 18). The Chief Economist of NASA 'has never once seen us use a CBA as an important tool to make an argument' (email to AO, 12 Oct. 2022). Much of the data required is only available retrospectively, all the more so in the pursuit of the unknown. A CBA surplus is neither necessary nor a sufficient condition for projects like DES.

In CBA future benefits (and future costs) are discounted on the commonsense assumption that a dollar today is worth more than one in a year's time. In competitive markets the discount rate is the market interest rate which serves as a benchmark for profit-making (Offer 2022: 18, 32-33, 158-160; Price 1993: ch. 8). Market discount rates are high and applying such rates restricts projects to those which can pay off in a short period of time. For long-term endeavours such as astrophysics it is not clear that discounting is appropriate at all since payoffs are so nebulous and remote (Mishan 2002: 16). In practice discounting appears to be ignored. In any case the components of cost-benefit analysis are so inexact that a low discount rate implies a spurious precision. For example, the value of the early career researchers' lifetime earnings premium, which is notionally the largest social payoff, is very sensitive to the discount rate, and is unlikely to be known to decision makers with any precision, though it might be estimated retrospectively.

In CBA willingness to pay is regarded as a valid determinant of social value. This can be elicited in surveys, aggregated, and compared with the cost. If the total is greater than the cost, that can be taken as a good reason to proceed. But this signal is far from crisp. Its magnitude can be estimated by measuring tax expenditures on science per taxpayer. The tax outlays on the CERN Hadron Collider are on the order of a few euro a year per taxpayer, and in surveys, the stated willingness to pay is more than the actual tax burden in France, and substantially more in Switzerland (Florio 2019: 250; Florio 2023). That give rise to a puzzle. On the one hand it is assumed that scientific expenditure is delegated knowingly to decision-makers by informed citizens, that it is undertaken with consent from 'ordinary people' (Florio 2019: 254) and confers a benefit upon them. To some extent, the value of science depends on what government is willing to pay for it. That leaves little room to estimate value added, since the outlay is the same as the cost: there is no financial surplus. It could be argued however that the surplus of individual taxpayers' willingness to pay over and above what is actually paid in tax represents the social value added. As against that argument, this willingness is just an average, with substantial numbers expressing little or no appreciation for scientific expenditure and no desire to support it. Furthermore, while economists assume that people know what is good for them, there is a some skepticism about this even among CBA authorities. Individual choice is subject to many biases. For example, individuals commonly discount the future inconsistently, steeply at the outset, much less so for distant payoffs. They show a strong myopic bias. That is at odds with the standard procedure of discounting in CBA, which remains consistent throughout (Brent 2017: 80-85). An untested willingness to pay by the average taxpayer can hardly provide the motivation for expensive scientific projects. If it is the willingness to pay of the funding agency that is being measured, one has to presume that the agency is expecting some return in excess of what they pay, but what they pay is the only thing we know.

It might be worth thinking of the net benefits of scientific discovery as a 'quasi-option' value, i.e. as the willingness to pay for an uncertain and yet unknown discovery or to defer a decision in order to avoid being locked prematurely into a course of action (Florio 2019: ch. 9). For example, DES has the option value of detecting an empirical anomaly to Einstein's general relativity. It is correct however that there is no certainty that the flow of benefits is going to be positive. If nuclear war happens, any survivors may regard the splitting of the atom as unfortunate after all.

The CBA approach only captures what can be measured. 'The actual meaning… is to focus on what can be said quantitatively, with due caution, compared to what can be said only qualitatively' (Florio, 2019: 285). That is a bias in itself given that much of the payoff cannot be precisely quantified. Given however that cost-benefit valuation is either speculative, or requires retrospective data, the surplus of estimated benefit over cost cannot in itself be said to explain the motivation for research infrastructures, or for projects like the DES. The motive to invest needs to be sought elsewhere.

For state funders, there is no measurable value-added benefit. The benefit is represented by the cost. CBA results provide a sense of what further benefits to look for and that perhaps is their main justification: it may be instructive here to list the main social benefits identified by Florio's CBA for CERN's Large Hadron Collider. In total, the excess of social benefit over cost is estimated with many caveats at about 21 percent (Florio 2019: 287-290). The benefits themselves break down as follows (table 1):

Table 1. CERN Large Hadron Collider Share of Total value-added by Category

| Category | Share |
| --- | --- |
| Scientific publications (citations) | 2% |
| Human capital formation | 33% |
| Technological spillovers | 32% |
| Cultural effects | 13% |
| Public good ('Existence value') | 20% |

*Source:* Florio (2019): 289.

Three experienced economists of science (David et al. 1992) have rejected altogether the validity of CBA in science, mostly on the ground that substantial economic downstream effects are generated but are difficult to trace. Florio himself is also cautious: 'policymakers will never accept or reject an RI project based solely on its expected socio-economic impact' (Florio 2019: 284). So what is it that actually drives projects like DES? Of the items in table 1 the final two (amounting to one-third) are non-use items that assume that science has some prior value.

Policymakers draw on the legacy culture in which they were educated and out of which they have done well personally. This is the legacy of secondary and university education in the humanities, mathematics, and science, in what is known in Germany as 'Bildung' (Lepenies 2006). In North-Western Europe in particular, societies were governed by an administrative elite selected by competitive examination from among university graduates. These are the people who have most successfully internalised and are able to articulate the tradition in which they were educated. The pursuit of science and the arts draws on this legacy. Many of the key figures of 'Big Science' in the United States also drew upon this European tradition or came directly out of it. As elite education has shifted increasingly towards the social sciences, with their utilitarian and 'me-first' orientation, the influence of elite culture has become increasingly tenuous.

Another legacy is that of metaphysical quest, and of a large metaphysical public sector, namely organised religion, to which science has been added since the 19th century. Before the discovery and erection of telescopes the greatest material artifacts of this tradition were the cathedrals of Europe, constructed to probe the inscrutable heavens in accordance with theological preconceptions. Societies defer to priests and shamans who claim privileged insight into abstruse knowledge. Astrophysics is in keeping with that legacy.

Society invests in museums, libraries, universities, and organised religion. Eighteenth-century English landowners surrounded themselves with vast landscape gardens designed to block out any suggestion of productive activity (Floud 2019: chs. 3-4). Modern democracies have inherited this tradition to create both urban and national parks (Price 2017). On the face of it, astrophysics also belongs in this tradition of imaginative creativity. The primary value of astrophysics would seem to be intrinsic and not instrumental, i.e. as an end in itself, and not a means for something else. It is part of the purpose of being human. Telling stories about the night sky is part of every culture. Astrophysics is the application of Enlightenment reason to the structure of the heavens. Social institutions reflect a community's intrinsic preoccupations.

Space exploration began visually, and since the 17$^{th}$ century has been carried out by means of increasingly sophisticated and costly equipment. It was funded initially by European kings and princes, continued in the United States by capitalist benefactors and their foundations, and carried out within universities, and by a variety of private inventors and enthusiasts (MacDonald 2017). The recent low altitude space travel stunts of the billionaires Musk, Bezos and Branson fall within this tradition of private gratification.

**Astrophysics and artificial intelligence**

Yet another social role may also be emerging for astrophysics. During the last two decades the development of artificial intelligence (AI) has accelerated and is now being pursued intensively in the private and public sectors alike. Artificial intelligence confers a competitive advantage on those who can make it work for themselves in business, government, in international competition and in war. Hence it currently appears to be unstoppable. Workers in this field are largely agreed that Generalised Artificial Intelligence (GAI), indistinguishable from human intelligence, is likely to be achieved within a generation or so, and there are outlying sceptics in both directions (Ford 2018; Brockman 2020). The timescale hardly matters. 30 years and 300 years are both fairly short periods in historical terms, and the blink of an eye in biological, geological and cosmic ones. Humanity is currently struggling with intractable environmental and resource problems. The conjunction of AI and global warming trends has given rise to broad speculation among astrophysicists and AI investigators that intelligence might evolve to expand into post-human nonbiological computational embodiments, perhaps in networked forms. The Internet and the Google search engine might both be seen as precursors, with the smartphone in every pocket serving as local nodes. Civilisation is only a few thousand years old, very short in evolutionary terms, let alone cosmic ones. If intelligence detaches itself from carbon dependency and biological habitats it can migrate into the galaxy and into the universe. Some astrophysicists and computer scientists already hold out this vision (Rees 2018: 150-153, 163; Tegmark 2017: ch. 6). From that perspective astrophysics may be regarded as a preliminary reconnaissance for the migration of intelligence into stellar spaces. Dark energy comes into it too. It tells us that in an expanding universe even an intelligence unbounded by biology can only colonise a small fraction of the whole. Astrophysics may yet map out the next great step for reason and consciousness. The ultimate payoffs may be stupendous but CBA will be irrelevant.

**Conclusion**

For its funders DES is a model enterprise. But for all its brio, in scientific terms it was less than fully satisfying. Its resident anthropologist explains why: 'Every scientist I spoke to would be happy if DES data infer that the current best prediction for the dark energy parameter ($w=-1$) is incorrect.' (Calder and Holbraad 2021: 353). 'Professor M' (actually one of us, the current author OL) is quoted there as saying, 'It would be nice if it makes a landmark discovery… whether or not we'll find something new depends on whether nature is kind enough to unveil it' (Ibid. 351). But nature has not been kind. In Popperian terms there is no falsification. In terms of Kuhn's model of scientific revolutions DES is puzzle solving normal science, which disappointingly delivers what it promises. The quest for novelty has ended up with a precise confirmation. But that may be the wrong way to think about it. Such precise confirmation by empirical means remains a work of wonder which highlights the sublime genius of Einstein's General Relativity. Astronomy is also a form of exploration, and the DES continues the work of Columbus and Magellan by mapping new domains, collecting data, cataloguing, expanding the known universe, laying down a hoard of data and placing it in the public domain. In this respect it was always bound to succeed and is a justified source of pride. It is in the eyes of the beholder if the DES finding of agreeing with the standard LCDM is 'good' or 'bad' news. The role of the experiment is to discover, as precisely and accurately as possible, what is out there in nature. The interpretation of results is a different matter.

This study does not have a breakthrough finding. The social value of dark energy is more diffuse than an economist would like. Precise estimates are not reliable, and reliable estimates are not precise. But it is better to be vaguely right than precisely wrong. It is useful to know that like much of human enterprise astrophysics is undertaken without being signed off by accountants. At the present state of knowledge that may be our foremost finding, and a challenge for further research. The best determinant of how much is going to be done is how much was done last year and the year before. Periodic assessments in the form of grant allocations are much more narrowly focused than the spectrum of benefits identified here. We know approximately how much astrophysics costs and much less precisely what benefits it provides. In the aggregate, that is enough to maintain the quest at its current level. Astrophysics has few goods to sell. It maintains the incentive structure of the historical 'republic of science' due to its long-term, open-ended nature. The reality of the starry sky is not a human construct, however much its interpretation may be. Theory in astrophysics is in constant interaction with observation and measurement, and there is far less scope to structure it for private benefit. In a world of fake news, malign ideologies, and propaganda, astrophysics, like art and science more generally, retains its human appeal. It is part of the quest for security which is embodied in impartial, disinterested truth.

**Declarations – Competing interests**

No funds, grants or other support was received. The authors have no relevant financial or non-financial interests to disclose.

**Declarations – Compliance with ethical standards**